\documentclass[bjps,preprint]{imsart}
\usepackage{amsfonts}
\usepackage{graphicx}
\usepackage{amsmath}
\usepackage{amssymb}%
\usepackage{float}
\providecommand{\U}[1]{\protect\rule{.1in}{.1in}}
\newtheorem {theorem}{Theorem}[section]

\newtheorem{lemma}{Lemma}[section]

\newtheorem{remark}{Remark}[section]

\newcommand{\E}{\mathbb{E}}

\newcommand{\bi}[1]{\mbox{\boldmath{$ #1 $}}} 
\begin{document}

\begin{frontmatter}

\title{A rank-based Cram{\' e}r-von-Mises-type test for two samples}
\runtitle{A rank based two-sample test}

\begin{aug}
\author{\fnms{Jamye} \snm{Curry}\thanksref{a}\ead[label=e1]{jcurry4@ggc.edu}},
\author{\fnms{Xin} \snm{Dang}\thanksref{b,t1}\ead[label=e2]{xdang@olemiss.edu}}
\and
\author{\fnms{Hailin} \snm{Sang}\thanksref{b}%
\ead[label=e3]{sang@olemiss.edu}}%

\thankstext{t1}{Corresponding author}

\affiliation[a]{Georgia Gwinnett College}
\affiliation[b]{University of Mississippi}

\address{Address of Jamye Curry \\School of Science \& Technology,\\ Georgia Gwinnett College, \\Lawrenceville, GA 30043, USA.\\
\printead{e1}}

\address{Address of Xin Dang and Hailin Sang\\Department of Mathematics, \\University of Mississippi,\\
University, MS 38677, USA.\\
\printead{e2,e3}}

\end{aug}

\begin{abstract}
We study a rank based univariate two-sample distribution-free test. The test statistic is the difference between the average of between-group rank distances and the average of within-group rank distances. This test statistic is closely related to the two-sample Cram{\' e}r-von Mises criterion. They are different empirical versions of a same quantity for testing the equality of two population distributions. Although they may be different for finite samples, they share the same expected value, variance and asymptotic properties. The advantage of the new rank based test over the classical one is its ease to generalize to the multivariate case. Rather than using the empirical process approach, we provide a different easier proof, bringing in a different perspective and insight. In particular, we apply the H{\' a}jek projection and orthogonal decomposition technique in deriving the asymptotics of the proposed rank based statistic. A numerical study compares power performance of the rank formulation test with other commonly-used nonparametric tests and recommendations on those tests are provided. Lastly, we propose a multivariate extension of the test based on the spatial rank. 

\end{abstract}



\end{frontmatter}

\section{Introduction}
To test whether two samples come from the same or different populations,  several distribution free tests such as the Kolmogorov-Smirnov test, the Cram{\' e}r-von Mises test and their variations have been proposed and widely used. Let $X_1, X_2,..., X_m  \stackrel{iid}{\sim} F$ and $Y_1, Y_2,..., Y_n\stackrel{iid}{\sim} G$  be two independent random samples with continuous distribution functions $F$ and $G$, respectively.  The two sample problem is to test
\begin{equation}\label{hypothesis}
H_0: F=G\;\;\;vs\;\;\; H_a: F\neq G.
\end{equation}
Denote $F_m$ and $G_n$ as the empirical distribution functions of the two samples and $H_N$  as the empirical distribution function of the combined sample, where $N=m+n$. The Kolmogorov-Smirnov (KS) two-sample test uses the maximum distance (difference) between $F_m$ and $G_n$. The classical Cram{\' e}r-von Mises test statistic has the form
\begin{align}\label{CM}
T_{c}=\frac{mn}{N}\int_{-\infty}^\infty[F_m(x)-G_n(x)]^2dH_N(x).
\end{align}
This test statistic and its asymptotics have been well studied in the literature, for example, Lehmann \cite{Lehmann51}, Rosenblatt \cite{Rosenblatt52}, Darling \cite{Darling57}, Fisz \cite{Fisz60} and Anderson \cite{Anderson62}.

Both of the KS test statistic and the Cram{\' e}r-von Mises test statistic are formulated based on the empirical distributions. Sz{\'e}kely and Rizzo \cite{Szekely04}, Baringhaus and Franz \cite{Baringhaus04} studied a test statistic based on the original data. That is
\begin{align}\label{eqn:dist}
\frac{mn}{N}\{\frac{1}{mn}\sum_{i=1}^m\sum_{j=1}^n |X_i-Y_j|
 -\frac{1}{2m^2}\sum_{i=1}^m\sum_{j=1}^m|X_i-X_j|
 -\frac{1}{2n^2}\sum_{i=1}^n\sum_{j=1}^n |Y_i-Y_j|\}.
\end{align}
This test has a direct generalization to the multivariate case. However, it requires an assumption on the first moment and it is not distribution free for the univariate case.  It is worth to note that the test statistic \eqref{eqn:dist} falls in the unified framework on energy statistics studied by Sz{\' e}kely and Rizzo~\cite{Szekely13, Szekely17} and can be easily generalized to the $K$ sample problem. Other similar tests include  \cite{Fernandez08} and \cite{Gretton08}, although they are derived under different motivations.  Fernandez, Gamerro, and Garc\`{i}a \cite{Fernandez08} developed a statistic based on the empirical characteristic functions of the observed observations.  The statistic uses a weighted integral of the difference between the empirical characteristic function of the two samples.  Gretton et al. \cite{Gretton08} proposed a test based on a kernel method in which the testing procedure is defined as the maximum difference in expectations over functions evaluated on the two samples. All of those test statistics are of the form being a difference on a measure of between-group and within-group.  

In this paper, we propose a new rank based test of the same form. Nevertheless, it overcomes the limitations of (\ref{eqn:dist}). It is formulated based on the ranks of two samples with respect to the combined sample $H_N$.  Denote $R(y, H)$ as the standardized rank of the quantity $y$ with respect to the distribution $H$, i.e., $R(y,H)=H(y)$.
For testing the hypothesis (\ref{hypothesis}), we use the following test statistic.
\begin{align}\label{eqn:statistic1}
 T=&\frac{mn}{N}\{\frac{1}{mn}\sum_{i=1}^m\sum_{j=1}^n |R(X_i,H_N)-R(Y_j,H_N)|\nonumber \\
 &-\frac{1}{2m^2}\sum_{i=1}^m\sum_{j=1}^m|R(X_i,H_N)-R(X_j,H_N)| \nonumber \\
 &-\frac{1}{2n^2}\sum_{i=1}^n\sum_{j=1}^n |R(Y_i,H_N)-R(Y_j,H_N)|\}.
\end{align}
$T$ is interpreted as the difference of the average of between-group rank differences and the average of within-group rank differences.  A large value of $T$ indicates the deviation of two groups. The test based on $T$ is distribution-free and does not require any moment condition. 

For the balanced samples ($m=n$), one can consider an equivalent but simpler statistic 
\begin{eqnarray}\label{eqn:statisticT}
T^\prime= \frac{1}{mn}\sum_{i=1}^m\sum_{j=1}^n |R(X_i,H_N)-R(Y_j,H_N)|.
\end{eqnarray}
$T^\prime$ is the average of rank differences between two groups. $T$ and $T^\prime$ are equivalent because $T = nT^\prime -(4n^2-1)/(12n)$ when $m=n$. 

As we will see later, the test statistic $T$ is closely related to the classical nonparametric Cram{\' e}r-von Mises criterion $T_c$. They are different empirical plug-in versions of the same population quantity. The rank based test statistic and the Cram{\' e}r-von Mises criterion may not be equal to each other for finite samples, but they are asymptotically equivalent. The advantage of the new rank based test over the classical one is its ease to generalize to the multivariate case. Multivariate generalizations of Cram{\' e}r-von Mises tests have been considered by many researchers, but they are either applied on independent data \cite{Cotterill82} or used for testing independence \cite{Genest07} or used in the goodness-of-fit test of the uniform distribution on the transformed data \cite{Chiu09}. For the rank based formulation, generalizations to the multivariate two sample problem are straightforward by applying notions of multivariate rank functions.  In this paper, rather than using the empirical process approach, we provide a different easier proof, bringing in a different perspective and insight. In particular, we apply the H{\' a}jek projection and orthogonal decomposition technique in deriving the asymptotics of the proposed statistic.

Some related works include Pettitt \cite{Pettitt76} and Baumgartner, Wei\ss,  and Schindler \cite{Baumgartner98}. They considered statistics of Anderson-Darling type that can be viewed as standardized versions of Cram{\' e}r-von Mises statistics. Schmid and Trede \cite{Schmid95} utilized $\mathcal{L}^1$ Cram{\' e}r-von Mises statistics. A rank-based representation of a $\mathcal{L}^1$ Cram{\' e}r-von Mises statistic under a balanced size and its generalizations are studied by Borroni \cite{Borroni01}. Albers, Kallenberg and Martini \cite{Albers01} studied rank procedures for detecting shift alternatives with increasing shift in the tail of the distribution. Janic-Wr\'{o}blewska and Ledwina \cite{Janic00} considered a test based on a combination of several linear rank statistics. Related to the rank procedures, other nonparametric tests include those based on the empirical likelihood approach. Einmahl and McKeague \cite{EM03} considered test statistics based on the empirical likelihood ratios for the goodness of fit and two sample problems. It has been proved that those tests are asymptotically equivalent to the one-sample and two sample Anderson-Darling tests. Cao and Van Keilegom \cite{CK06}  proposed an empirical likelihood ratio test via kernel density estimation. Gurevich and Vexler \cite{GV11} utilized an empirical likelihood ratio test based on samples entropy.   

The paper has the following structure. Section 2 presents the main results, including the formulation of the test statistic and its properties.  The simulation study is performed in Section \ref{sim}. We propose a multivariate extension of the test in Section \ref{mul}. We summarize and conclude the paper in Section \ref{sd}. All proofs go to Section \ref{proof}.

\section{Main Results}

To formulate the rank based test statistic $T$ in (\ref{eqn:statistic1}), we first establish its population version. We provide a result of the population version, from which we can see the relationship between our statistic and Cram{\' e}r-von Mises criterion. 

\begin{theorem}\label{thm:inequality}
Let $X,  X_1, X_2$ and $Y,  Y_1, Y_2$ be independent continuous random variables distributed from $F$ and $G$, respectively. Let $H=\tau F+(1-\tau)G$ with $0\leq\tau\leq 1$ be the mixture distribution. Then
\begin{equation} \label{eqn:inequality}
 \E|R(X,H)-R(Y,H)|-\frac{1}{2}\E|R(X_1,H)-R(X_2,H)|-\frac{1}{2}\E|R(Y_1,H)-R(Y_2,H)|\geq0
\end{equation}
and the equality holds if and only if $F=G$.
\end{theorem}
The above result is based on the following identity which
is obtained from Lemma \ref{lemma:int} in the Appendix.
\begin{align}
&\E |R(X,H)-R(Y,H)|-\frac{1}{2}\E|R(X_1,H)-R(X_2,H)|-\frac{1}{2}\E|R(Y_1,H)-R(Y_2,H)|  \nonumber \\
&=\int_{-\infty}^\infty (F(x)-G(x))^2d(\tau F(x)+(1-\tau)G(x)).  \label{FG}
\end{align}

The result of Theorem \ref{thm:inequality} suggests two possible statistics for testing the hypothesis (\ref{hypothesis}). The first version is the sample plug-in version of the left side of \eqref{FG}.  With $\tau=m/N$ and multiplying by $mn/N$, it is our test statistic defined in (\ref{eqn:statistic1}).  $H_0$ is rejected if the sample version is large, i.e., $T > c_{\alpha}(m,n)$. The critical value $c_{\alpha}(m,n)$ is determined by the significance level $\alpha$ and the null distribution of $T$. The test statistic $T$ is the difference of the average of between-group rank differences and the average of within-group rank differences. A large value of $T$ indicates the deviation of two groups.

The two-sample Cram{\' e}r-von Mises statistic $T_c$ in (\ref{CM}) is the empirical version of the right side of (\ref{FG}).  Hence  $T$ and  $T_c$ are all plug-in statistics of an equal quantity. Nevertheless, they may take different values. We shall thank one of the referees who pointed out this possibility. For example, in the case that $m=n=2$, let the two $X$ realizations be $0$ and $2$ and the  two $Y$ realizations be $1$ and $3$. It is easy to see that the Cram{\' e}r-von Mises statistic has value $\frac{1}{4}$ and the test statistic $T$ has value $\frac{1}{8}$. Next, we will study the properties of $T$. 

Let $D$ be $\E |R(X,H)-R(Y,H)|-\frac{1}{2}\E|R(X_1,H)-R(X_2,H)|-\frac{1}{2}\E|R(Y_1,H)-R(Y_2,H)|$, and  $\hat D = N/(mn) T$. Then we have the following theorem. 
\begin{theorem}\label{SLLN}
For $m,  n \rightarrow \infty$,  if  $m/(m+n) \rightarrow \tau$, then $\hat{D}\rightarrow D\;\;a.s.$
\end{theorem}
By this theorem and Theorem \ref{thm:inequality}, it is easy to see that our test statistic $T$ is consistent for the alternative $H_a: F\neq G$. 

\begin{theorem}\label{free}
Under $H_0$, $T$ is distribution free.
\end{theorem}

Under $H_0$, the combined samples $X_1,...,X_m, Y_1,...,Y_n$ constitute a random sample of size $N$ from the distribution $F=G=H$. So any assignment of $m$ numbers to $X_1,...,X_m$ and $n$ numbers to $Y_1,...,Y_n$ from the set of integers $\{1,2,...,N\}$ is equally likely, i.e. has probability $\dbinom{N}{m}^{-1}$ which is independent of $F$. Using the fact that those number assignments have one-to-one linear relationship with the standardized ranks, $T$  is distribution free.

\begin{figure}[tbh]
\centering
\begin{tabular}{cc}
\includegraphics[width=0.5\linewidth]{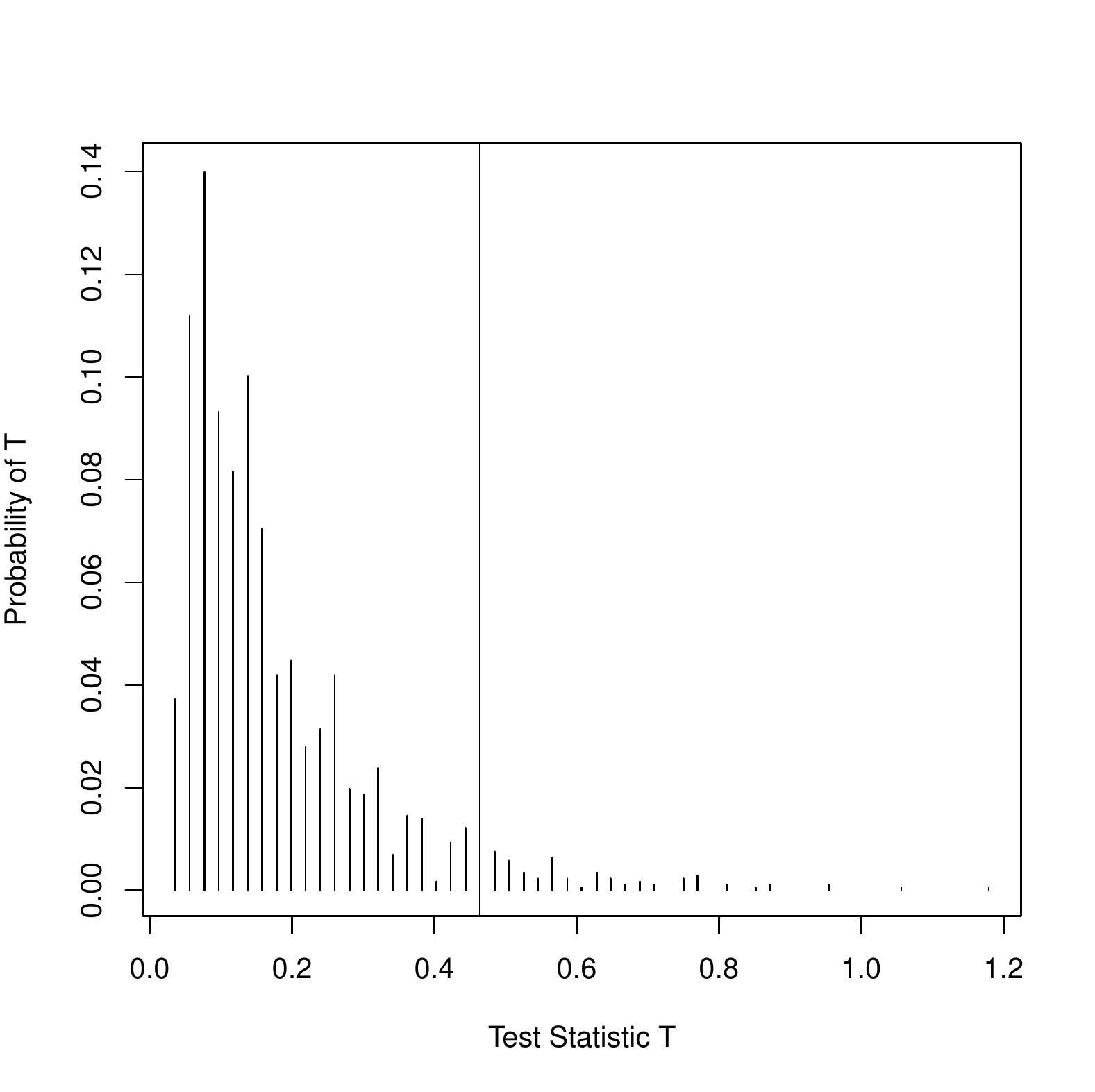}
&
\includegraphics[width=0.5\linewidth]{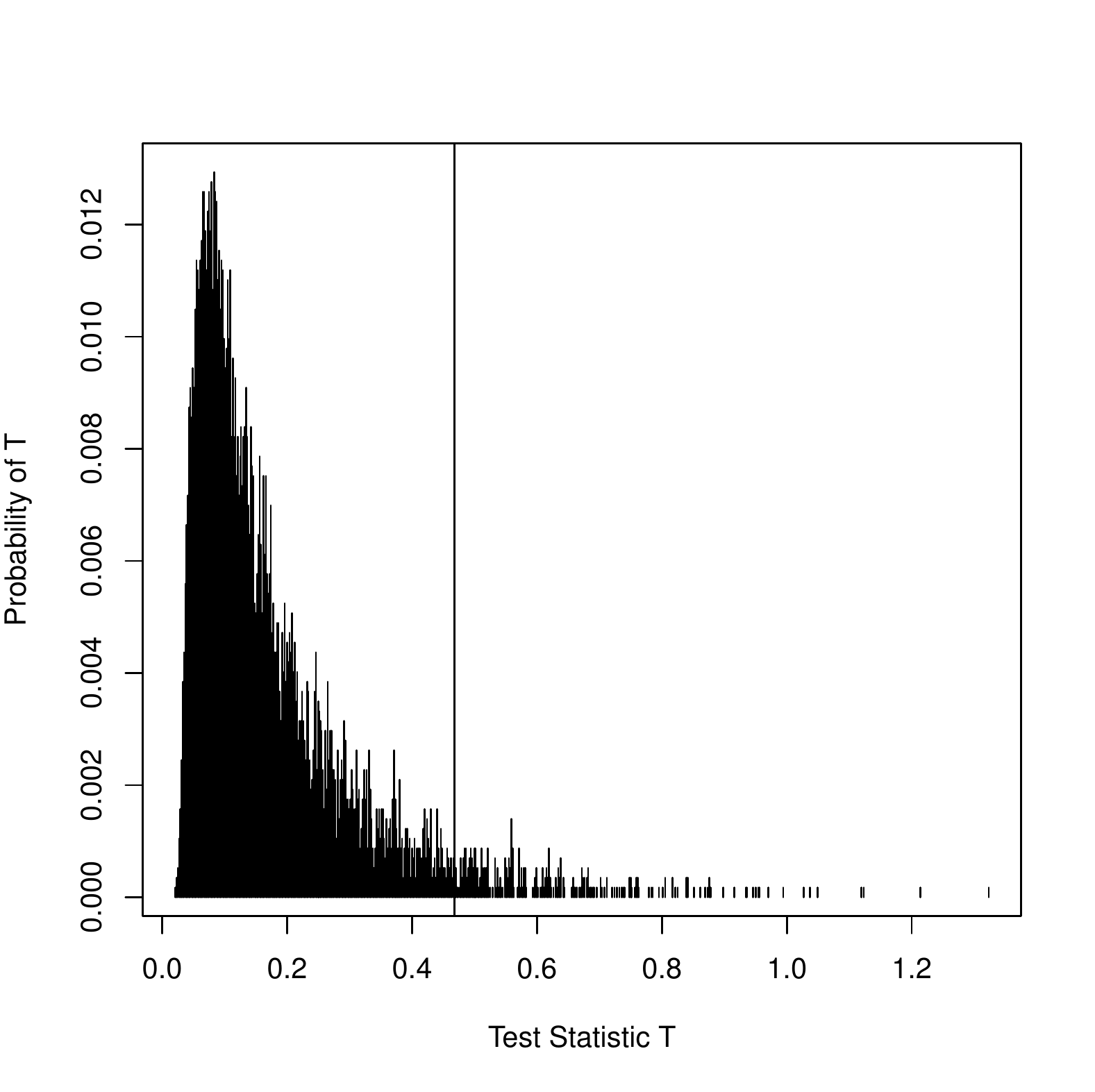}
\end{tabular}
\caption {The exact null distribution of $T$ obtained from all combinations (Left: $m=7$, $n=7$; Right: $m=7$, $n=9$). The vertical line in each graph indicates the 5\% critical value. \label{fig:exactdist}}
\end{figure}

The exact null distribution of $T$ can be found by enumeration of all possible values of $T$  by considering the $N!/(m! n!)$ orderings of $m$ $X$'s and $n$ $Y$'s.  Figure \ref{fig:exactdist} provides the exact null distribution of $T$ for sample sizes $m=n=7$ and $m=7, n=9$ by considering all combinations.  However, the exact null distribution is infeasible to obtain for large sample sizes because the number of combinations increases dramatically as $m$ and $n$ increase. For large samples, we can use Monte-Carlo method on all combinations to approximate the null distribution. Also the limiting distribution of $T$ can be used to determine the critical values of the test.  Next we study asymptotic behaviors of $T$.   

Since $T$ and the Cram{\' e}r-von Mises statistic $T_c$ are different sample plug-in versions of a same quantity for checking the equality of two distributions, we expect that they should have the same  expectation, variance and asymptotic distribution under the null hypothesis.  The expectation and variance of $T_c$ are provided by Anderson \cite{Anderson62}.  In the next theorem, we obtain the same results for $T$, but provide a more straightforward  derivation and simpler proof.   
\begin{theorem}\label{t1var}
Under $H_0$, $$\E T=\frac{N+1}{6N}=\frac{1}{6}+\frac{1}{6N}$$ and
$$Var(T)=\frac{N+1}{180N^2}\left[4(N+1)-\frac{3N^2}{mn}\right].$$
\end{theorem}
\begin{remark}
In particular, if $m, n\rightarrow \infty$, $\E T\rightarrow 1/6$ and $Var(T)\rightarrow 1/45$.
\end{remark}

Rosenblatt \cite{Rosenblatt52} and Fisz \cite{Fisz60} have derived the limiting distribution of Cram{\' e}r-von Mises statistic, which is a mixture of independent $\chi^2_1$ distributions. It is necessary to check whether or not our test statistic has the same limiting distribution. Rather than using a stochastic process method, we provide a different H{\' a}jek projection approach to obtain the limiting distribution of $T$, which agrees with that of $T_c$.    

We obtain the first order H{\' a}jek projection $\tilde T$ in Lemma \ref{lemma3.1} as  
 \begin{eqnarray}
 \tilde T=\sum_{i=1}^m\E[T|X_i]+\sum_{j=1}^n\E[T|Y_j]-(N-1)\E T,\nonumber
 \end{eqnarray}
where $\E[T|X_1]=\frac{1}{6}(1+\frac{n}{mN})+\frac{1}{N}(1-\frac{n}{m})[F(X_1)-F^2(X_1)]$. 
Under $H_0$, $\tilde T$ has variance
 \begin{eqnarray}
 Var(\tilde{T}) &=&\sum_{i=1}^mVar(\E[T|X_i])+\sum_{j=1}^nVar(\E[T|Y_j])\nonumber\\
 &=&\left[\frac{m(m-n)^2}{m^2N^2}+\frac{n(m-n)^2}{n^2N^2}\right] Var[F(Y_1)-F^2(Y_1)]\nonumber\\
 &=&\frac{(m-n)^2}{180mnN}. \nonumber
 \end{eqnarray}
Clearly, $Var(\tilde{T})/Var(T)\rightarrow 0$ as $m, n\rightarrow \infty$. Therefore the  first order H{\' a}jek projection is not sufficient in deriving the asymptotics of the statistic $T$.   

To derive the asymptotics of the statistic $T$ under the null hypothesis, it is necessary to have the second order  projection $\hat{T}$ of $T$.
\begin{align*}
\hat{T}=&\sum_{i=1}^m\sum_{j=1}^n\E[T|X_i, Y_j]+\sum_{1\le i<j\le m}\E[T|X_i, X_j]\\
&+\sum_{1\le i<j\le n}\E[T|Y_i, Y_j]-\frac{N(N-1)}{2}\E T.
\end{align*}
Since 
\begin{align*}
&\E\{\E[T|X_1, Y_1]\}=\E\{\E[T|X_1, X_2]\}=\E\{\E[T|Y_1, Y_2]\}=\E T, 
\end{align*}
$\E \hat T=0$.
 By Lemma \ref{tx1y1}  and Lemma \ref{tx1x2},  it can be examined that 
$$Cov(\E[T|Z_1, Z_2], \E[T|Z_1, Z_3])=0,$$ 
where $Z_1, Z_2$ and $Z_3$ are three different variables from $X_i, Y_j$, $1\le i\le m, 1\le j \le n$, and 
\begin{align*}
Var(\hat{T})&=\frac{m^4+n^4-2m^3n-2mn^3+10m^2n^2-8mnN+5n^2+5m^2}{180N^2mn}\\
&+\frac{m(m-1)}{2}\frac{m^2-2mn+5n^2}{90m^2N^2}
+ \frac{n(n-1)}{2}\frac{n^2-2mn+5m^2}{90n^2N^2}\\
&=\frac{N^2}{180mn}+\frac{2}{45N}-\frac{N-1}{36mn}-\frac{1}{18N^2}.
\end{align*}
Then $\frac{Var(\hat{T})}{Var (T)}\rightarrow 1$ as $N\rightarrow \infty$ under the condition
$ \lim_{N\rightarrow \infty}m/n=1$. We shall always assume this condition in the following analysis. 
  
Efron and Stein \cite{EfronStein} discussed a general orthogonal decomposition of a statistic. Here, our statistic $T$ is decomposed as $\tilde T+\hat T+ R_N$, where $\tilde T$ is the first order projection and $R_N$ is a negligible term. Hence the limiting distribution of $T$ is determined by the limiting distribution of $\hat T$.

To determine the limiting distribution of $\hat T$ under $H_0$, let $h(x,y)=|F(x)-F(y)|+F(x)[1-F(x)]+F(y)[1-F(y)]-2/3$. Then $h(x,y)$ is a degenerate kernel function since $h(x,y)$ is symmetric and $\E h(X,y)=0$. By Lemma \ref{tx1y1} and Lemma \ref{tx1x2}, we have $\hat{T}=\hat{\hat{T}}+R_N'$ with $Var(R_N')/Var(T)\rightarrow 0$ as $N\rightarrow \infty$~\footnote{If $m=n\rightarrow \infty$, $R_N^{\prime}=0$.} and 
$$\hat{\hat{T}}=\frac{1}{N}\sum_{i=1}^m\sum_{j=1}^n h(X_i,Y_j)-\frac{1}{N}\sum_{1\le i<j\le m}h(X_i, X_j)-\frac{1}{N}\sum_{1\le i<j\le n}h(Y_i, Y_j).$$
It is not difficult to verify that
\begin{equation}\label{var}
Var [h(Z_1,Z_2)]=2/45
\end{equation}
and $Cov(h(Z_1, Z_2), h(Z_1, Z_3))=0$, i.e.,  $h(Z_1,Z_2)$ and $h(Z_1,Z_3)$ are orthogonal, where $Z_1, Z_2$ and $Z_3$ are three different variables from $X_i, Y_j$, $1\le i\le m, 1\le j\le n$. 

Now we define an operator $A$ on the function space $\mathcal{L}^2(\mathbb{R}, F)$ by
$$Ag^*(x)=\int_{-\infty}^\infty h(x,y)g^*(y)dF(y),\;\;\;x\in\mathbb{R}, \;g^*\in \mathcal{L}^2(\mathbb{R},F).$$
This operator only has real eigenvalues since the kernel $h(x,y)$ is symmetric. Let $\lambda=\lambda_1, \lambda_2, \cdots $ be the non-zero eigenvalues of the operator $A$ obtained by solving  the equation $Ag^*=\lambda g^*$.  With the substitution of $u=F(x)$ and $v=F(y)$, solving $Ag^*=\lambda g^*$ is equivalent to solve that
\begin{equation}\label{eigen}
 \int_0^1 \left\{|u-v|+u(1-u)+v(1-v)-\frac{2}{3}\right\}g(v)dv = \lambda g(u),
\end{equation}
where $g = g^*\circ F^{-1}$. Taking the twice derivative with respect to $u$ on both sides of (\ref{eigen}), we have the equation $2 g(u) = \lambda g^{\prime\prime}(u)$. Solving it and substituting back, we have the eigenvalues of $A$ being $\lambda_k=-\frac{2}{\pi^2k^2}, k\in \mathbb{N}$ and the corresponding eigenfunctions $\phi_k(x)=\cos (k\pi F(x)), k\in \mathbb{N}$. The eigenfunction for the zero eigenvalue is $\phi_0(x)=1$. Note that the eigenvalues do not depend on $F$, but the eigenfunctions $\{\phi_k(x)\}_{k=0}^\infty$ depend on $F$, which give a orthonormal basis for the space $\mathcal{L}^2(\mathbb{R},F)$. Let $T_N=\hat{\hat{T}}/\sqrt{Var(T)}$. Then we have the following theorem. 
\begin{theorem}\label{proasym}
Under $H_0$ and the condition $ \lim_{N\rightarrow \infty}m/n=1$, 
$$T_N \stackrel{d}{\longrightarrow} Z_{\infty}=-\frac{\sqrt{45}}{2}\sum_{k=1}^\infty \lambda_k(\chi_{1k}^2-1),$$
where $\chi_{11}^2,\chi_{12}^2\cdots$ are independent $\chi_{1}^2$ variables and $\lambda_k=-\frac{2}{\pi^2 k^2}, k\in \mathbb{N}$.
Hence
$$(T-\E T)/\sqrt{Var(T)}\stackrel{d}{\longrightarrow} Z_{\infty}=-\frac{\sqrt{45}}{2}\sum_{k=1}^\infty \lambda_k(\chi_{1k}^2-1)$$ since $(T-\E T-\hat{\hat{T}})/\sqrt{Var(T)}\rightarrow 0$ in probability.
\end{theorem}
As expected, this asymptotical result agrees with the one for Cram{\' e}r-von Mises statistic as proved with a stochastic process method in Rosenblatt \cite{Rosenblatt52} and Fisz \cite{Fisz60}. This different projection approach we applied here is typically useful in U-statistic theory, but we shall emphasize that  $T$ is not an U-statistic.

\begin{table}[thb] 
\center
\begin{tabular}{lrrrrr} \hline\hline
& $d=1$&$d=2$&$d=4$&$d=10$&$d=100$\\ \hline
Variance ratio& 0.9239&0.9819&0.9967&0.9997&1.0000\\
95\% quantile & 1.9298& 1.9676& 1.9772 &1.9779& 1.9780\\ \hline
\multicolumn{6}{l}{Approximated $c_{\alpha}(m,n)$ of $T$ ($\alpha=0.05$) based on $Z_d$}\\ \hline
 $m=n=50$&0.4545&0.4601& 0.4617& 0.4617 &0.4617\\
 $m=50$, $n=40$&0.4545&0.4601& 0.4615& 0.4616 &0.4616\\
 $m=n=500$&0.4544& 0.4600& 0.4614& 0.4615& 0.4615\\
 $m=n=7$ &0.4543& 0.4597&0.4610&0.4611& 0.4611\\
 $m=7$, $n=9$&0.4540& 0.4594& 0.4608& 0.4609& 0.4609\\ \hline\hline
\end{tabular}
\caption{First part: variance ratios of $Z_d$ over $Z_{\infty}$ and  $95\%$ quantiles of $Z_d$. Second part: approximated critical values for $T$. Comparing with the exact $\alpha=0.049$ critical value 0.4643  for the case of $m=n=7$  and the exact $\alpha=0.05$ critical value 0.4678 for the case of $m=7, n=9$, the approximations are pretty accurate even under small sizes. In practice, $d=4$ or $d=10$ is recommended. \label{dmix}}
\end{table} 

In practice, we may approximate the limiting distribution by a distribution of a finite linear combination of $d$ independent $\chi^2_1$ random variables, i.e.  
$$Z_d=-\frac{\sqrt{45}}{2}\sum_{k=1}^d \lambda_k(\chi_{1k}^2-1)= \frac{\sqrt{45}}{\pi^2}\sum_{k=1}^d \frac{1}{k^2}(\chi_{1k}^2-1).$$ The accuracy of approximation depends on the choice of $d$. Table \ref{dmix}  provides ratios of variance of the $d$ mixture and that of the infinite mixture, that is, $\sigma^2({Z_d})/\sigma^2(Z_{\infty})$. Also the table lists 95\% quantiles of $Z_d$ which are estimated by the average of 10 sample quantiles each on $M=10^8$ random samples. Those quantile values can be used to approximate the critical values $c_{\alpha}(m,n)$ of $T$, which are given by the second part of Table \ref{dmix}.  As we will see that even for small sample sizes, the approximated critical values are pretty accurate and close to the exact true values. For the case of $m=n=7$, the true size of the test is 0.056 if the approximated critical value 0.4611 is used. For the case of $m=7$ and $n=9$, the true size of the test is 0.052 if 0.4609 is used.  In summary, $d=4$ or $d=10$ is recommended for a compromise between computation and accuracy.

\section{Simulations}\label{sim}
By the simulation study in this section we demonstrate the performance of the T test.  There are many nonparametric tests available for the two sample problem. It is by no means to conduct a comprehensive comparison.  Here we include Kolmogorov-Smirnov test (KS), Wilcoxon rank sum test (W) or Mood test (M), the empirical likelihood ratio test (ELR) proposed by Gurevich anf Vexler \cite{GV11},  the empirical likelihood test (ELT) proposed by Einmahl and McKeague \cite{EM03}, Baringhaus and Franz's Cram\'{e}r test (CT),  the test studied in Fern\'{a}ndes et al. \cite{Fernandez08} (DT) in the study.  It is necessary to note that the CT and DT tests are not  distribution-free tests, and their critical values and p-values are based Monte-Carlo method on permutations in each sample, which is implemented in the R package ``cramer". The R package ``dbEmpLikeGOF" is used for the ELR test in which the parameter is set to be 0.1 as suggested in \cite{GV11}. The critical values of the ELT and our T test are computed through $10^7$ random combinations on $\{1, ..., N\}$.  

\begin{table}[thb]
\centering
\begin{tabular}{cccccccc} \hline\hline
$\Delta$ &           KS   &  W   & ELR   &ELT   &CT  & DT  &T \\ \hline 
0& 0.040& 0.050& 0.057& 0.050 &0.050& 0.051& 0.050\\  \vspace{0.2cm}
&0.041 &0.047 &0.031 &0.047& 0.048& 0.048 &0.049\\
0.25& 0.162& 0.228& 0.182& 0.224& 0.226& 0.191& 0.217\\\vspace{0.2cm}
&0.160& 0.208 &0.119 &0.196 &0.203 &0.171& 0.198\\
0.5& 0.534& 0.681& 0.578& 0.670& 0.671& 0.582& 0.652\\ \vspace{0.2cm}
&0.498& 0.621 &0.446 &0.603& 0.615& 0.526 &0.600\\
0.75& 0.875& 0.949&0.902& 0.945& 0.943& 0.901& 0.936\\\vspace{0.2cm}
&0.851& 0.926 &0.829 &0.919 &0.922& 0.871& 0.912\\
1& 0.988& 0.998& 0.994& 0.997& 0.998 &0.991& 0.996\\
&0.979 &0.995& 0.976& 0.994 &0.994& 0.984 &0.993\\ \hline\hline
\end{tabular}
\caption{Power performance of each test with significance level $\alpha=0.05$ for the normal distribution with location alternatives. Row 1: $n=m=50$, Row 2: $n=50, m=40$} \label{tab:norm}
\end{table}  

Various alternative distributions are considered.  For each case, $M=10000$ iterations are computed to estimate powers by calculating the fraction of p-values less than or equal to $\alpha =0.05$ the level of significance. The Monte Carlo errors can be estimated by $\pm 1.96 \sqrt{p(1-p)/M}$. In particular,  the size of tests shall maintain in the interval (0.046, 0.054).
\begin{table}[thb]
\centering
\begin{tabular}{cccccccc} \hline\hline
$\Delta$ &           KS   &  W   & ELR   &ELT   &CT  & DT  &T \\ \hline 
0& 0.036 &0.048& 0.052& 0.048 &0.049 &0.046& 0.047\\  \vspace{0.2cm}
&0.045 &0.054 &0.036 &0.054 &0.055 &0.051& 0.054\\
0.25& 0.130&0.163& 0.124& 0.160& 0.158& 0.142 &0.165\\\vspace{0.2cm}
&0.135 &0.157& 0.084& 0.152& 0.154& 0.137& 0.162\\
0.5& 0.422 &0.488 &0.367& 0.486 &0.481& 0.434 &0.501\\ \vspace{0.2cm}
&0.402 &0.449 &0.268& 0.440& 0.439& 0.394& 0.462\\
0.75&0.776& 0.830& 0.710 &0.825& 0.818& 0.783& 0.836\\\vspace{0.2cm}
&0.741& 0.786 &0.586 &0.780& 0.776& 0.734 &0.799\\
1& 0.947& 0.966& 0.917& 0.966& 0.965& 0.950 &0.971\\
&0.929 &0.946& 0.842 &0.946 &0.944& 0.925& 0.954\\ \hline\hline
\end{tabular}
\caption{Power performance of each test with significance level $\alpha=0.05$ for the $t_3$ with location alternatives. Row 1: $n=m=50$, Row 2: $n=50, m=40$.} \label{tab:t3}
\end{table}
Table \ref{tab:norm} shows the size and power performance for each test under the normal distributions, where $X_1$,$\dots$,$X_n$ $\sim N(0, 1)$ and $Y_1$,$\dots$,$Y_m$ $\sim N(\Delta, 1)$ with $\Delta=$ 0, 0.25, 0.5, 0.75, and 1.   When $\Delta=0$, the KS test is undersized for both the equal and unequal sample sizes cases; the ELR test is oversized in the equal sample size case and seriously undersized for the sample unequal size case; all other tests keep a desirable size.  As expected, the W test is the best among all tests since it is well-known to be powerful for the two-sample problem with a constant shift in location, especially when data follow logistic or normal distributions. The CT and ELT tests are comparable to W.  The T test is more powerful than the DT, KS and ELR tests. In the unequal sample size case, the W test is the best followed by the CT test. The ELT and T tests are comparable and significantly better than the DT, KS and ELR tests.
 
The experiment is repeated for the $t$-distribution with 3 degrees of freedom and  the result is presented in Table \ref{tab:t3}.   Although the statistical power of the $T$ test is the highest among all tests for all cases,  its power differences with the W test or the ELT test are small so that those three tests are comparable. 
\begin{table}[thb]
\centering
\begin{tabular}{cccccccc} \hline\hline
$\Delta$ &           KS   &  W   & ELR   &ELT   &CT  & DT  &T \\ \hline 
     0& 0.040& 0.052& 0.058& 0.051& 0.052& 0.050& 0.051\\  \vspace{0.2cm}
       & 0.044& 0.050& 0.032& 0.050& 0.050& 0.053& 0.052\\
0.25& 0.417& 0.443& 0.906& 0.599& 0.205& 0.287& 0.472\\\vspace{0.2cm}
       & 0.377& 0.405& 0.815& 0.516& 0.186& 0.265& 0.428\\
  0.5& 0.960& 0.886& 0.999& 0.978& 0.655& 0.843& 0.968\\ \vspace{0.2cm}
       & 0.940& 0.859& 0.998& 0.958& 0.598& 0.798& 0.948\\
0.75& 0.999& 0.989& 1.000& 0.999& 0.945& 0.993& 0.999\\\vspace{0.2cm}
       & 0.998& 0.980& 1.000& 0.998& 0.908& 0.988& 0.997\\
     1& 1.000& 0.999& 1.000& 1.000& 0.993& 1.000& 1.000\\ 
       & 1.000& 0.998& 1.000& 1.000& 0.988& 1.000& 1.000\\ 
\hline\hline
\end{tabular}
\caption{Power performance of each test with significance level $\alpha=0.05$ for the Pareto distributions with location alternatives. Row 1: $n=m=50$, Row 2: $n=50, m=40$.} \label{tab:pareto}
\end{table}
Table \ref{tab:pareto} shows the power performance for the Pareto distribution, where $X_1$,$\dots$,$X_n$ $\sim$ Pa(2, 2) and $Y_1$,$\dots$,$Y_m$ $\sim$ Pa(2+$\Delta$, 2) are generated, with $\Delta=$ 0, 0.25, 0.5, 0.75, and 1.    The power of the ELR test is much higher than that of all others.  For $\Delta =0.25$, the power of the ELR test  is as high as 90\%, which is 30\% higher than the second best  ELT test.  The T test is the third best one. The power difference between the $T$ test and that of the $CT$ test can be as large as 27\% for equal sample sizes and can be as large as 32\% for unequal sample sizes. 

All considered tests as in the experiment for location alternatives are used for scale alternatives except the Wilcoxon test (W), as this  is a test for location.  Instead, the Mood's test known as a scale test is used and referred to as the M test.  Table \ref{scale1} displays the results when $Y$ samples of size 50 are generated from $N(0,\Delta)$ or Pareto$(2, 2\Delta)$,  where $\Delta=$ 1, 1.5, 2, 2.5, and 3.  In the normal case, the T test does not compare favorably to all considered tests other than the KS test. It performs significantly better than the KS test, but its power is 2-5 times smaller than that of others.  It is interesting to see that the M test outperforms all tests in the normal case but it is the inferior in the Pareto case.  The T test has better performance for Pareto samples than for normal samples due to the heavy tails in Pareto distributions. In the Pareto case, all tests outperform the M test by a great margin and the CT test is the superior.  As suggested by a reviewer, we add one more case in the simulation in which $X_1,...,X_n \sim exp(1)$ and $Y_1,...,Y_m \sim lognorm(0,1)$ with sample sizes $m=n=50$.  The Monte Carlo powers of the seven tests are listed in Table \ref{scale1}.  In this scenario, the T test performs better than KS and DT, but does not compare as favorably to the CT, W, ELT and ELR tests. 
 \begin{table}[thb] 
\centering
 \begin{tabular}{lcccccccc} \hline\hline
Distribution &$\Delta$ &           KS   &  M   & ELR   &ELT   &CT  & DT  &T \\ \hline 
&1 &0.039& 0.051& 0.056& 0.047& 0.049& 0.051& 0.047\\
&1.5 &0.118& 0.663& 0.542& 0.238& 0.251& 0.431& 0.138\\
Normal&2 &0.374& 0.979& 0.965& 0.746& 0.792& 0.915& 0.479\\
&2.5 &0.681& 0.999& 0.999& 0.962& 0.981& 0.994& 0.815\\
&3&0.881& 1.000& 1.000& 0.996& 0.999& 1.000& 0.957\\ \hline
&1 &0.040& 0.054& 0.055& 0.049& 0.052& 0.049& 0.050\\
&1.5 &0.307& 0.098& 0.378& 0.418& 0.487& 0.356& 0.398\\
Pareto&2 &0.741& 0.165& 0.828& 0.857& 0.909& 0.815& 0.831\\
&2.5 &0.937& 0.214& 0.973& 0.980& 0.992& 0.974& 0.978\\
&3 &0.988& 0.234& 0.997& 0.998& 0.999& 0.997& 0.998\\ \hline
\multicolumn{2}{l}{Exp vs Lgnorm} & 0.336 &0.535 &0.654 & 0.555& 0.502& 0.315 &0.476\\
 \hline\hline
\end{tabular}
\caption{Power performance of each test with significance level $\alpha=0.05$ for Normal and Pareto scale alternatives, also the case of F = Exp and G = Lognorm. \label{scale1} }
\end{table}

In general, the T test is not recommended for scale alternatives. The Kolmogorov-Smirov test is not recommended either. The empirical likelihood ELR test is more suitable for a general scale alternative, but is not recommended for a location alternative for symmetric distributions. The T test has a better performance for location alternatives than scale alternatives.  It is easy to explain the power performance of the Cram{\' e}r-von Mises test with the rank based formulation (\ref{eqn:statistic1}) for the location alternatives. For two samples from the same class distributions (normal distributions, t distributions or Pareto distributions and so on) but with different locations,  the ranks in the mixture are quite different. Therefore the corresponding test can easily recognize them and have good power performance. We recommend to apply the T test for location alternatives, especially in the heavy-tailed distributions.


\section{Multivariate Extension}\label{mul}
The proposed rank test statistic is closely related to the two sample Cram{\' e}r-von Mises criterion. Both statistics are different sample plug-in forms from a same population quantity. The advantage of our rank test is to allow straightforward generalizations to the multivariate case by using different multivariate rank functions. Among them, the spatial rank is appealing due to its computation ease, efficiency and other nice properties \cite{Mottonen97}, \cite{Oja10}. The sample version of the spatial rank function with respect to $H_N$, the empirical distribution of the combined sample $\bi x_1,...,\bi x_m$ and $\bi y_1,...,\bi y_n$  in $\mathbb R^d$,  is defined as 
$$\bi R(\bi x, H_N)=\frac{1}{N}\sum_{i=1}^N \frac{\bi x-\bi z_i }{\|\bi x-\bi z_i\|}, $$ where $\bi z_i = \bi x_i$ for $i=1,...,m$, $\bi z_{m+i}=\bi y_i$ for $i=1,...,n$ and $\|\cdot\|$ is the Euclidian distance. Then the  multivariate two-sample spatial rank statistic, denoted by $T_{M}$, is defined as
\begin{eqnarray}
T_{M}
&=&\frac{mn}{N}\{\frac{1}{mn}\sum_{i=1}^m\sum_{j=1}^n \|\bi R(\bi x_i,H_N)-\bi  R(\bi y_j,H_N)\|  \nonumber\\
&& - \displaystyle\frac{1}{2m^2}\sum_{i=1}^m\sum_{j=1}^m\|\bi R(\bi x_i,H_N)-\bi R(\bi x_j,H_N)\| \nonumber\\
&& - \displaystyle\frac{1}{2n^2}\sum_{i=1}^n\sum_{j=1}^n \|\bi R(\bi y_i,H_N)-\bi R(\bi y_j,H_N)\|\}. \label{eqn:TM}
\end{eqnarray}
The test statistic $T_{M}$ is the difference of the average of the intra-group rank distances and the average of the inter-group rank distances. A large value of $T_{M}$ indicates the deviation of the two groups and rejects the null hypothesis.  The multivariate counterpart of Theorem \ref{thm:inequality} states as follows. 

\begin{theorem} \label{thm:multi-inequality}
Let $\bi X,  \bi X_1, \bi X_2$ and $\bi Y,  \bi Y_1,\bi Y_2$ be independent $d$-variate continuous random vectors distributed from $F$ and $G$, respectively. Let $H=\tau F+(1-\tau)G$ with $0\leq\tau\leq 1$. Then
\begin{align} 
&\E\|\bi R(\bi X,H)-\bi R(\bi Y,H)\|-\frac{1}{2}\E\| \bi R(\bi X_1,H)-\bi R(\bi X_2,H)\| \nonumber\\ 
&-\frac{1}{2}\E\|\bi R(\bi Y_1,H)-\bi R(\bi Y_2,H)\| \geq0, \label{eqn:mveqn}
\end{align}
where the equality holds if and only if $F=G$. 
\end{theorem}

The multivariate spatial rank test based on $T_M$ loses the distribution-free property under the null hypothesis.  The test relies on the permutation method to determine critical values or compute p-values. But the test is robust. For example, it does not require the assumption of finite second moment as the Hotelling's $T^2$ test. Neither it requires the assumption of finite first moment as the test (CT) considered by Baringhaus and Franz \cite{Baringhaus04}.  

 A simulation is conducted to compare performance of $T_M$, CT and the Hotelling's $T^2$ under  multivariate normal, $t_1$ and Pareto distributions on $\mathbb R^d$ ($d=2, 5$). Location and scatter alternatives are considered. For location alternatives in normal and $t_1$ distributions,  the parameters of distributions for generating $\bi X$ samples of size $n=50$ are $\bi \mu = \bi 0$ and $\bi \Sigma_X = \bi I$, while for $\bi Y$ samples with size $m=50$ are $\bi \mu = (\Delta,...,\Delta)^T$ and $\bi \Sigma_Y=\bi I$, where $\Delta= 0,0.25,0.5,0.75$ and 1.  For Pareto distribution, $\bi X = (X_1,...,X_d)^T$ is generated with each component $X_j$ from Pareto(1,1) and $\bi Y =(Y_1,...,Y_d)^T$ is generated with each component $Y_j$ from Pareto$(1+\Delta,1)$.  R package ``Hotelling" is used for the Hotelling's $T^2$ test. $T_M$ and CT tests use the permutation method to compute p-values and $M=10000$ iterations are computed to estimate powers by calculating the fraction of $p$-values less than or equal 0.05. Results for the location alternatives are listed in Table \ref{tab:table7}. 
 
 \begin{table}[thb]
\centering
\begin{tabular}{l c c|ccccc} \hline\hline
Dist & Dim & Method & $\Delta=0$& $\Delta=.25$&$\Delta=.50$&$\Delta=.75$&$\Delta=1$\\ \hline
&&$T_M$ &  0.0550 & 0.3000& 0.8688 & 0.9966  &  1\\
&$d=2$&CT & 0.0556  &0.3090& 0.8818  &0.9972 &   1\\
&& Hotelling & 0.0518&  0.3226& 0.8900&  0.9976&    1\\ \cline{2-8}
Norm
& &$T_M$& 0.0484&  0.5178& 0.9944&  1&    1\\
&$d=5$&CT&0.0500 & 0.5332 &0.9958&  1&    1\\
&& Hotelling & 0.0494& 0.5248&0.9942 & 1   & 1\\ \cline{2-8}
  \hline
& &$T_M$&0.0538 & 0.1574& 0.4898 & 0.8212& 0.9644 \\
&$d=2$& CT & 0.0596 & 0.0820 & 0.226&  0.4504& 0.7134\\
&& Hotelling &  0.0546 & 0.0562 &0.0934&   0.1360& 0.2058\\ \cline{2-8}
$t_1$
& &$T_M$& 0.0478 & 0.2382& 0.7986&  0.9888& 0.9996\\
&$d=5$&CT&0.0546 & 0.0858&  0.288&  0.6200& 0.8568\\
&& Hotelling & 0.0472 & 0.0742 &0.1622&   0.2990& 0.4608\\ \cline{2-8}
  \hline
&&$T_M$&0.0492 & 0.3470 &0.8682 & 0.9886& 0.9998 \\
& $d=2$& CT &0.0560 & 0.1146& 0.2850  &0.5330 &0.7298\\
&& Hotelling &  0.0484&  0.0986 &0.1858 & 0.3076& 0.4188\\ \cline{2-8}
Pareto
& &$T_M$& 0.0522  &0.2892& 0.7942&  0.9784& 0.9996\\
   &$d=5$&CT&0.0492&  0.1142& 0.2942 & 0.5184& 0.7128\\
      && Hotelling & 0.0528&  0.1108 &0.2614&  0.4462& 0.6046\\
 \hline \hline
\end{tabular}
\caption{Power performance of $T_M$, CT and Hotelling tests with significance level $\alpha=0.05$ for multivariate normal, $t_1$ and Pareto distributions with location alternatives with sample sizes $n=m=50$.} \label{tab:table7}
\end{table}

From Table \ref{tab:table7}, three tests keep the size 5\% well. Powers in $d=5$ are higher than that in $d=2$ for each of three tests under all distributions.  In the normal cases, $T_M$ performs slightly worse than the Hotelling's test and CT. The power of $T_M$ is about $2\%$ lower than that of the Hotelling test and $1\%$ lower than that of CT under $H_a$ when $\Delta =0.25$ and $\Delta=0.50$.  However, the power gain of $T_M$ over CT and the Hotelling's test is huge in the $t_1$-distributions. For $\Delta=0.25$ and $\Delta=0.5$, $T_M$ is about twice powerful as CT and about triple powerful as the Hotelling test. The advantage of our proposed $T_M$ over CT and the Hotelling's test are even more significant in the asymmetric Parato distributions than in the $t_1$ distributions for the location alternatives.   

\begin{table}[thb]
\centering
\begin{tabular}{l c c|ccccc|r} \hline\hline
Dist & Dim & Method & $\Delta=1$& $\Delta=1.5$&$\Delta=2$&$\Delta=2.5$&$\Delta=3$& Orient\\ \hline
&&$T_M$ &    0.0468&    0.0640&  0.1064&    0.1902&  0.2992&0.3179\\
&$d=2$&CT & 0.0474&    0.0982&  0.2716&    0.5660&  0.8072& 0.3016\\
&& Hotelling &  0.048 &   0.0472 & 0.0598 &   0.0496 & 0.0524&0.0493\\ \cline{2-9}
Norm
& &$T_M$&  0.0476&    0.0748&  0.1272 &   0.2600&  0.4124&0.9678\\
&$d=5$&CT& 0.0470&    0.1510&  0.5580&    0.9192&  0.9948&0.8188\\
&& Hotelling &   0.045&    0.0568 & 0.0526&    0.0576&  0.0540&0.0538\\ \cline{2-8}
  \hline
& &$T_M$&0.0486 &   0.0580 & 0.0698&    0.0948&  0.1256&0.2366\\
&$d=2$& CT &  0.0482 &   0.0680  &0.1182    &0.1754  &0.2370&0.0916\\
&& Hotelling & 0.0506 &   0.0476&  0.0488    &0.0530&  0.0544& 0.0495\\ \cline{2-9}
$t_1$
& &$T_M$&0.0514&    0.0648  &0.0900&    0.1286 & 0.1742&0.5896\\
&$d=5$&CT&0.0512 &   0.0836&  0.1510&    0.2320&  0.3344&0.1984\\
&& Hotelling &  0.0528&    0.0494 & 0.0492&    0.0556 & 0.0550&0.0468\\ \cline{2-9}
  \hline
&&$T_M$& 0.0550 &   0.6164&  0.9802&     1 &      1&-\\
& $d=2$& CT &0.0540&    0.6896&  0.9896&         1&       1&-\\
&& Hotelling & 0.0498&    0.5148&  0.9354    &0.9876&  0.9976&-\\ \cline{2-9}
Pareto
& &$T_M$&  0.0504 &   0.9158&  1   &  1&       1&-\\
   &$d=5$&CT&0.0512 &   0.9268&  0.9996&         1&       1&-\\
      && Hotelling & 0.0566   & 0.7616&  0.9960   & 0.9998&  1&-\\
 \hline \hline
\end{tabular}
\caption{Power performance of $T_M$, CT and Hotelling tests with significance level $\alpha=0.05$ for multivariate normal, $t_1$ and Pareto distributions with Scatter alternatives with sample sizes $n=m=50$.} \label{tab:table8}
\end{table}
Results for scatter alternatives are listed in Table \ref{tab:table8}. For multivariate normal and $t_1$ distributions,  we first consider the difference of scatter matrix only on scales. The parameters for $\bf X$ sample are $\bi \mu = \bi 0$ and $\bi \Sigma_X = \bi I$, while for $\bi Y$ samples  are $\bi \mu =\bi 0$ and   and $\bi \Sigma_Y=\Delta \bi I$, where $\Delta=1,1.5,2,2.5$ and 3.    We then consider the alternative with different orientation on the scatter matrices. The scatter matrix is $\left(\begin{array}{lr} 1&.5\\.5&1\end{array} \right)$ for $\bf X$ samples, while it is $\left(\begin{array}{lr} 1&-.5\\-.5&1\end{array} \right)$ for $\bf Y$ samples. Hence two components of $\bf X$ are positively correlated and the two components of $\bf Y$ are negatively correlated. The results for orientation difference alternatives are listed in the last column "Orient" of Table \ref{tab:table8}.  In $d=5$, $\bi \Sigma_X$ has diagonal elements to be 1 and off-diagonal elements to be 0.5 and $\bi \Sigma_Y$ is constructed to have the same eigenvectors as $\bi \Sigma_X$ and eigenvalues to be the reciprocals of eigenvalues of $\bi \Sigma_X$.  For Pareto distributions, $\bi X = (X_1,...,X_d)^T$ is generated with each component $X_j$ from Pareto(1,1) and $\bi Y =(Y_1,...,Y_d)^T$ is generated with each component $Y_j$ from Pareto$(1,\Delta)$.

From Table \ref{tab:table8}, all tests maintain the size 5\% well. For asymmetric Pareto distributions, CT is slightly better than $T_M$ and $T_M$ is better than the Hotelling's test. For normal and $t_1$ distributions, the Hotelling's $T^2$ completely fails in scatter alternatives since it is a test on location difference.  CT test is much better than $T_M$ for scale alternatives. Particularly CT is triple powerful as the $T_M$ in normal case and twice  powerful in the $t_1$ case.   This result is not surprising since $T_M$ is based on the spatial ranks that lose major information on distances or scales.  However, when two scatter matrices of distributions are different on orientation, $T_M$ performs better than CT, especially in $t_1$ distribution, the power of $T_M$ is twice or triple as that of CT. 

\section{Summary}\label{sd}

The problem of testing whether two samples come from the same or different population is a classical one in statistics. In this paper, we have studied a rank-based test for the univariate two sample problem. The test statistic is the difference between the average of between-group rank distances and the average of within-group rank distances. Under the null hypothesis, it is distribution free.  The limiting null distribution was explored through techniques of H{\' a}jek projection and orthogonal decomposition. It has been proved that the limiting distribution is not normal since the projection on one variable is insufficient to represent the variation of the test statistic. By taking the second-order projection, an operator in the functional space was defined and its eigenfunctions and eigenvalues were applied to derive the limiting distribution.  It is a weighted mixture of independent chi-square distributions with the weights being the eigenvalues of the operator. We provided a recommendation how to use the limiting distribution to obtain critical values of the proposed test in practice. 

The proposed rank test statistic is closely related to two sample Cram{\' e}r-von Mises criterion. Both statistics are different sample plug-in forms from the same population quantity. We have provided a counter example to show they are different. However, they have the same expectation, variance and limiting distribution. The advantage of our rank test is to allow straightforward generalizations to the multivariate case by using different multivariate rank functions.  A continuation of this work is to study properties of the multivariate Cram{\' e}r-von Mises $T_M$ test. Also the generalizations based on other multivariate rank functions deserve further investigation.

\section{Proofs}\label{proof}
The following lemma gives the expected value of the absolute difference between the standardized ranks of $X$ and $Y$. 
\begin{lemma}\label{lemma:int}
Let $X$ and $Y$ be independent continuous random variables from $F$ and $G$, respectively.  Let $H=\tau F+(1-\tau)G$ with $0\leq\tau\leq 1$ be the mixture distribution, $J$ be the distribution of $R(X, H)$ and $K$ be the distribution function of $R(Y,H)$. Then
\begin{equation}\label{expd}
\mathbb{E} |R(X,H)-R(Y,H)|=\int_{0}^{1} J(t)(1-K(t))\,dt +\int_{0}^1K(t)(1-J(t))\,dt.
\end{equation}
\end{lemma}
\noindent{\bf Proof of Lemma \ref{lemma:int}}. Notice that 
\begin{align*}
&|R(X,H)-R(Y,H)|\\
&=\int_{0}^{1}[I(R(X,H)\leq s<R(Y,H))+I(R(Y,H)\leq s<R(X,H))]\,ds.
\end{align*}
Since $H$ is continuous and $R(X, H)=H(X), R(Y, H)=H(Y)$, we have $J(x)=F\circ H^{-1}(x)$, $K(x)=G\circ H^{-1}(x)$ for any $x\in[0, 1]$, where $H^{-1}(x)=\inf \{u: H(u)\ge x\}$. Then (\ref{expd}) holds by Fubini's Theorem.    \hfill{$\square$}
\\

\noindent{\bf Proof of Theorem \ref{SLLN}}. Define 
\begin{align*}
 \tilde{D}=&\frac{1}{mn}\sum_{i=1}^m\sum_{j=1}^n |R(X_i,H)-R(Y_j,H)|\nonumber \\
 &-\frac{1}{2m^2}\sum_{i=1}^m\sum_{j=1}^m|R(X_i,H)-R(X_j,H)| \nonumber \\
 &-\frac{1}{2n^2}\sum_{i=1}^n\sum_{j=1}^n |R(Y_i,H)-R(Y_j,H)|.
\end{align*}
Conditioning on $Y_j$, $1\le j\le n$, by the law of large numbers, 
\begin{align*}
\frac{1}{m}\sum_{i=1}^m |R(X_i,H)-R(Y_j,H)|-\E_{X_1} |R(X_1,H)-R(Y_j,H)|\rightarrow 0 \;\;a.s., 
\end{align*}
and 
\begin{align*}
&\frac{1}{mn}\sum_{i=1}^m\sum_{j=1}^n |R(X_i,H)-R(Y_j,H)|\\
&=\frac{1}{n}\sum_{j=1}^n \left[\E_{X_1} |R(X_1,H)-R(Y_j,H)|+o_{a.s.}(1)\right]\\
&=\E |R(X,H)-R(Y,H)|+o_{a.s.}(1). 
\end{align*}
By the strong law of large numbers for $U$-statistics \cite{Hoeffding61}, 
\begin{align*}
\frac{1}{2m^2}\sum_{i=1}^m\sum_{j=1}^m|R(X_i,H)-R(X_j,H)|\rightarrow \frac{1}{2}\E|R(X_1,H)-R(X_2,H)|\;\;a.s., 
\end{align*}
and 
\begin{align*}
\frac{1}{2n^2}\sum_{i=1}^n\sum_{j=1}^n|R(Y_i,H)-R(Y_j,H)|\rightarrow \frac{1}{2}\E|R(Y_1,H)-R(Y_2,H)|\;\;a.s. 
\end{align*}
Then 
$ \tilde{D}\rightarrow D \;\; a.s.$ as $m,n\rightarrow \infty$. Now we show $\hat{D}-\tilde{D}\rightarrow 0\;\;a.s.$\\
By Glivenko-Cantelli theorem, for $m/N \rightarrow \tau$, 
\begin{align*}
R(x,H_N)\rightarrow R(x,H)\;\;a.s.
\end{align*}
uniformly on $x\in \mathbb{R}$. Then 
\begin{align*}
|R(x,H_N)-R(y,H_N)|\rightarrow |R(x,H)-R(y,H)|\;\;a.s.
\end{align*}
uniformly on $x, y\in \mathbb{R}$ and therefore,
\begin{align*}
\hat{D}-\tilde{D}=&\frac{1}{mn}\sum_{i=1}^m\sum_{j=1}^n  \left\{|R(X_i,H_N)-R(Y_j,H_N)|- |R(X_i,H)-R(Y_j,H)|\right\}\nonumber \\
 &-\frac{1}{2m^2}\sum_{i=1}^m\sum_{j=1}^m \left\{|R(X_i,H_N)-R(X_j,H_N)|-|R(X_i,H)-R(X_j,H)|\right\} \nonumber \\
 &-\frac{1}{2n^2}\sum_{i=1}^n\sum_{j=1}^n \left\{|R(Y_i,H_N)-R(Y_j,H_N)|-|R(Y_i,H)-R(Y_j,H)|\right\}.\\
& \rightarrow 0\;\;a.s.
\end{align*}
This completes the proof.  \hfill{$\square$}

\vspace{4mm}

Since all sample ranks are computed with respect to the combined sample, for simple presentation, we write the standardized rank $R(X_i,H_N)$ as  $R(X_i)$. We also denote the natural rank of $X_i$ as $R_i$, that is, $R(X_i)=R_i/N$.


\vspace{0.2cm}
\noindent{\bf Proof of Theorem \ref{t1var}}. Under $H_0$, $R(X_i)$ and $R(Y_j)$ are identically distributed from the discrete uniform distribution on $\{1/N,2/N,...,(N-1)/N,1\}$ for all $i$'s and $j$'s. Hence
\begin{align*}
&\E T\\
=&\frac{mn}{N}\left\{\E |R(X_1)-R(Y_1)|
-\frac{m-1}{2m}\E|R(X_1)-R(X_2)|-\frac{n-1}{2n}\E|R(Y_1)-R(Y_2)|\right\} \nonumber\\
=&\frac{1}{2}\E|R(X_1)-R(X_2)|=\frac{1}{2N}\E|R_1-R_2|, \nonumber
\end{align*}
where $R_1$ and $R_2$ are natural ranks of $X_1$ and $X_2$, respectively. Under $H_0$, $R_1$ and $R_2$ are two samples drawn uniformly from $\{1,2,...,N\}$ without replacement. Clearly,
\begin{align}\label{eqn:r1r2}
\E|R_1-R_2|=\frac{1}{N(N-1)}\sum_{i=1}^N\sum_{j=1}^N |i-j|=\frac{N+1}{3}.
\end{align}
Hence $\E T=\displaystyle\frac{N+1}{6N}$.

Let $R_1, R_2, R_3, R_4$ be natural ranks of $X_1,X_2,X_3$ and $X_4$, respectively.  We have
\begin{align}
&\E (R_1-R_2)^2=\displaystyle\frac{1}{N(N-1)}\sum_{i=1}^N\sum_{j=1}^N(i-j)^2=\frac{N(N+1)}{6}, \label{e1}\\
&\E|R_1-R_2||R_1-R_3|=\frac{1}{N(N-1)(N-2)}\sum_{i=1}^N\sum_{j\neq i}\sum_{k\neq i,j}|i-j||i-k| \nonumber\\
&=\frac{1}{N(N-1)(N-2)}\sum_{i=1}^N\sum_{j=1}^N\sum_{k\neq j}|i-j||i-k|\nonumber\\
&=\frac{(N+1)(7N+4)}{60},\label{e2}\\
&\E|R_1-R_2||R_3-R_4| \nonumber\\
&=\frac{1}{N(N-1)(N-2)(N-3)}\sum_{i=1}^N\sum_{j\neq i,j=1}^N\sum_{k\neq i,j,k=1}^N\sum_{l\neq i,j,k; l=1}^N|i-j||k-l| \nonumber\\
&=\frac{1}{N(N-1)(N-2)(N-3)}\sum_{i=1}^N\sum_{j=1}^N\sum_{k\neq i,j,k=1}^N\sum_{l\neq i,j,k; l=1}^N|i-j||k-l| \nonumber\\
&=\frac{(N+1)(5N+4)}{45}. \label{e3}
\end{align}
Now extending $T^2$ and $\E T^2$ yields the above three types of expectations $\E(R_1-R_2)^2$, $\E |R_1-R_2||R_1-R_3|$ and $\E |R_1-R_2||R_3-R_4|$ denoted as $E_1,E_2, E_3$, respectively.  That is,
\begin{align*}
&\E T^2=\frac{m^2n^2}{N^4}\left\{\left[\frac{mn}{m^2n^2}+\frac{2m(m-1)}{4m^2}+\frac{2n(n-1)}{4n^4}\right] E_1\right.\\
&+\left[\frac{mn(m-1)(n-1)}{m^2n^2}+\frac{m(m-1)(m-2)(m-3)}{4m^4}+\frac{n(n-1)(n-2)(n-3)}{4n^4}\right.\\
&\left.-\frac{2mn(m-1)(m-2)}{2m^3n}-\frac{2mn(n-1)(n-2)}{2mn^3}+\frac{2m(m-1)n(n-1)}{4m^2n^2}\right] E_3\\
&+\left[\frac{m^2n^2-mn-mn(m-1)(n-1)}{m^2n^2}+\frac{4m(m-1)(m-2)}{4m^4}+\frac{4n(n-1)(n-2)}{4n^4}\right.\\
&\left.\left.-\frac{m^2n(m-1)-mn(m-1)(m-2)}{m^3n}-\frac{mn^2(n-1)-mn(n-1)(n-2)}{mn^3}\right]E_2\right\}\\
&=\frac{N+1}{60N^2}\left[3N+3-\frac{N^2}{mn}\right].
\end{align*}
Hence, 
\vspace*{-1cm}
\[
Var(T)=\E T^2-(\E T)^2=\frac{N+1}{180N^2}\left[4(N+1)-\frac{3N^2}{mn}\right]. \]  
\vspace*{-0.3cm}
This completes the proof. \hfill{$\square$}


\begin{lemma}\label{lemma3.1}
Under $H_0$, $\E [|R(X_1)-R(X_2)||X_1]=\displaystyle\frac{1}{2}-\frac{N-2}{N}[F(X_1)-F^2(X_1)]$,

 $\E[ |R(X_2)-R(X_3)||X_1]=\displaystyle\frac{1}{3}+\frac{2}{N}[F(X_1)-F^2(X_1)]$ and hence 

 $\E(T|X_1)=\displaystyle\frac{1}{6}\left(1+\frac{n}{mN}\right)+\frac{1}{N}\left(1-\frac{n}{m}\right)[F(X_1)-F^2(X_1)]$.
\end{lemma}
 
\noindent{\bf Proof of Lemma \ref{lemma3.1}}. Let $R_1$ be the natural rank of $X_1$ and $R_2$ be the natural rank of $X_2$. Under $H_0$ and given $X_1$, $R_1-1$  has a binomial distribution with parameters $N-1$ and $F(X_1)$, that is,
$$P(R_1-1=u|X_1)=\dbinom{N-1}{u}F(X_1)^u[1-F(X_1)]^{N-1-u}, \;\;\;u=0,1,...,N-1.$$
Given $R_1$, $R_2$ is uniformly distributed from $\{1,2,...,N\}/\{R_1\}$.  So
\begin{align*}
&\E(|R_1-R_2||X_1)\\
=&\E[\E(|R_1-R_2||R_1)|X_1] =\sum_{r_1=1}^N\frac{1}{N-1}\left[\sum_{i=1,\neq r_1}^N |r_1-i|\right]P(R_1=r_1|X_1)\\
=&\frac{1}{2(N-1)}\sum_{r_1=1}^N [(N-1)N-2(N-1)(r_1-1)+2(r_1-1)^2]P(R_1=r_1|X_1)\\
=&\frac{1}{2}N -(N-1)F(X_1)+\frac{1}{N-1}[(N-1)F(X_1)(1-F(X_1)+(N-1)^2F^2(X_1)]\\
=&\frac{1}{2}N-(N-2)[F(X_1)-F^2(X_1)].
\end{align*}
Let $R_3$ be the natural rank of $X_3$. Under $H_0$, we have
\begin{align*}
\E(|R_2-R_3||X_1)&= \E[\E(|R_2-R_3||R_1)|X_1]\\
&=\sum_{r_1=1}^N\frac{1}{(N-1)(N-2)}\left[\sum_{i=1,i\neq r_1}^N\sum_{j=1,j\neq i, r_1}^N |i-j|\right]P(R_1=r_1|X_1)\\
&=\frac{N(N+1)}{3(N-2)}-\frac{N-2(N-2)[F(X_1)-F^2(X_1)]}{N-2}\\
&=\frac{1}{3}N+2[F(X_1)-F^2(X_1)].\\
\end{align*}
It is clear that $\E(T|X_1)$ contains the above conditional expectations $\E[|R(X_1)-R(X_2)||X_1]$ and $\E[|R(X_2)-R(X_3)||X_1]$ denoted as $E_1^*$ and $E_2^*$, respectively. Then it follows that  
\begin{align*}
&\E (T|X_1)\\
&=\frac{mn}{N}\left\{\frac{n}{mn}E_1^*+\frac{mn-n}{mn}E_2^*-\frac{2(m-1)}{2m^2}E_1^*-\frac{(m-1)(m-2)}{2m^2}E_2^*-\frac{n(n-1)}{2n^2}E_2^*\right\}\\
&=\frac{mn}{N}\left\{ \frac{1}{m^2}(E_1^*-E_2^*)+\frac{N}{2mn}E_2^*\right\}\\
&=\frac{1}{6}\left(1+\frac{n}{mN}\right)+\frac{1}{N}\left(1-\frac{n}{m}\right)[F(X_1)-F^2(X_1)].
\end{align*}
This complete the proof.   \hfill{$\square$}

\vspace{4mm}

The following two lemmas are on the second order projection of $T$.
\begin{lemma}\label{tx1y1}
Under $H_0$, the second order projection of $T$ on one  $X$ variable and one $Y$ variable is
\begin{align}\label{conMxy}
&\E[T|X_1, Y_1]\nonumber\\
&=\frac{mnN+N^2-7mn}{6mnN}
+\frac{(2m-n)(n-1)}{mnN}F(X_1)[1-F(X_1)]\nonumber\\
&+\frac{(2n-m)(m-1)}{mnN}F(Y_1)[1-F(Y_1)]\nonumber\\
&+I(Y_1>X_1)\left\{\frac{5n-m}{6mnN}
+\frac{m-1}{mN}F(Y_1)-\frac{n-1}{nN}F(X_1)\right\}\nonumber\\
&+I(Y_1<X_1)\left\{\frac{5m-n}{6mnN}
-\frac{m-1}{mN}F(Y_1)+\frac{n-1}{nN}F(X_1)\right\}
\end{align}
and its variance is
\begin{align}
&Var \{\E[T|X_1, Y_1]\}\nonumber\\
&=\frac{m^4+n^4-2m^3n-2mn^3+10m^2n^2-8mnN+5n^2+5m^2}{180N^2m^2n^2}.\label{xyproj}
\end{align}
\end{lemma}

\noindent{\bf Proof of Lemma \ref{tx1y1}}. In the proof of next two lemmas, we use $\E[S|Z_1, Z_2, Z_1<Z_2]$ to denote $\E[SI(Z_1<Z_2)|Z_1, Z_2]$ for any random variables $S,  \;Z_1$ and $Z_2$. Again, let $R_i$ be the natural rank of $X_i$, $i=1,2,3, 4$. Under $H_0$ and given $X_1<X_2$, $(R_1, R_2)$ has trinomial distribution with parameters $F(X_1), F(X_2)-F(X_1)$ and $1-F(X_2)$, i.e.,
\begin{align}\label{r1r2}
&P(R_1=u, R_2=v|X_1, X_2, X_1<X_2)\\
&=\dbinom{N-2}{u-1, v-u-1,N-v}[F(X_1)]^{u-1}\\
&\times[F(X_2)-F(X_1)]^{v-u-1}[1-F(X_2)]^{N-v}I(X_1<X_2).\nonumber
\end{align}
Therefore,
\begin{equation}\label{1212}
\E[|R_1-R_2||X_1, X_2, X_1<X_2]=\{(N-2)[F(X_2)-F(X_1)]+1\}I(X_1<X_2).
\end{equation} 

Under $H_0$ and given $R_1,R_2$, the natural rank $R_3$ of $X_3$ has discrete uniform distribution on the set $\{1,2,\cdots, N\}/\{R_1, R_2\}$, i.e.,
 $$P(R_3=w|R_1, R_2)=\frac{1}{N-2}$$
 for $1\le w\le N$, $w\ne R_1, R_2$. Therefore,
 \begin{align}\label{1312}
 &\E(|R_1-R_3||X_1, X_2, X_1<X_2)\nonumber\\
 &=\E[\E(|R_1-R_3||R_1<R_2)|X_1, X_2, X_1<X_2]\nonumber\\
 &=\E\left[\frac{1}{N-2}\left(\sum_{1\le i<R_1}(R_1-i)+\sum_{R_1<i\le N, i\ne R_2}(i-R_1)\right)|X_1, X_2, X_1<X_2\right]\nonumber\\
  &=\E\left(\frac{N^2+N-2NR_1+2R_1^2-2R_2}{2(N-2)}|X_1, X_2, X_1<X_2\right)\nonumber\\
  &=\left\{\frac{N+1}{2}-(N-3)F(X_1)[1-F(X_1)]-F(X_2)\right\}I(X_1<X_2).
 \end{align}
 The last equality (\ref{1312}) is from (\ref{r1r2}), the conditional distribution of $(R_1, R_2)$. By a similar calculation,
  \begin{align}\label{1312'}
 &\E(|R_1-R_3||X_1, X_2, X_1>X_2)\nonumber\\
 &=\left\{\frac{N-1}{2}-(N-3)F(X_1)[1-F(X_1)]+F(X_2)\right\}I(X_1>X_2).
 \end{align}
 We also have
  \begin{align}\label{3412}
 &\E(|R_3-R_4||X_1, X_2, X_1<X_2)\nonumber\\
 &=\E[\E(|R_3-R_4||R_1<R_2)|X_1, X_2, X_1<X_2]\nonumber\\
 &=\E\left[\frac{1}{(N-2)(N-3)}\sum_{1\le i, j\le N, i,j\ne R_1, R_2}|i-j||X_1, X_2, X_1<X_2\right]\nonumber\\
  &=\left\{\frac{N-1}{3}+2F(X_1)[1-F(X_1)]+2F(X_2)[1-F(X_2)]\right\}I(X_1<X_2).
 \end{align}
 Again, the last equality (\ref{3412}) is from (\ref{r1r2}), the conditional distribution of $(R_1, R_2)$.
Now let $R_1$, $R_2$ be the natural rank of $X_1$ and $Y_1$ respectively. $R_3$ and $R_4$ be the natural ranks of two other different $X_i$ or $Y_j$, $1<i\le m, 1<j\le n$. By the definition of $T$  as in (\ref{eqn:statistic1}),
\begin{align*}
\frac{N}{mn}&\E[T|X_1, Y_1, X_1<Y_1]=\frac{1}{mn}\left\{\E[|R(X_1)-R(Y_1)||X_1, Y_1, X_1<Y_1]\right.\\
&+(n-1)\E(|R(X_1)-R(Y_2)||X_1, Y_1, X_1<Y_1)\\
&+(m-1)\E(|R(X_2)-R(Y_1)||X_1, Y_1,X_1<Y_1)\\
&+(m-1)(n-1)\E(|R(X_2)-R(Y_2)||X_1, Y_1, X_1<Y_1)\}\\
&-\frac{1}{2m^2}\left\{2(m-1)\E(|R(X_1)-R(X_2)||X_1, Y_1, X_1<Y_1)\right.\\
&\left.+(m-1)(m-2)\E(|R(X_2)-R(X_3)||X_1, Y_1, X_1<Y_1)\right\}\\
&-\frac{1}{2n^2}\left\{2(n-1)\E(|R(Y_1)-R(Y_2)||X_1, Y_1, X_1<Y_1)\right.\\
&\left.+(n-1)(n-2)\E(|R(Y_2)-R(Y_3)||X_1, Y_1, X_1<Y_1)\right\}\\
&=\frac{1}{mnN}\E(|R_1-R_2||X_1, Y_1, X_1<Y_1)\\
&+\frac{n-m}{m^2nN}\E(|R_1-R_3||X_1, Y_1, X_1<Y_1)\\
&+\frac{m-n}{mn^2N}\E(|R_2-R_3||X_1, Y_1, X_1<Y_1)\\
&+\frac{mnN+6mn-2N^2}{2m^2n^2N}\E(|R_3-R_4||X_1, Y_1, X_1<Y_1).
\end{align*}
By (\ref{1212}), (\ref{1312}), (\ref{1312'}) and (\ref{3412}), we have
\begin{align*}
&\E[T|X_1, Y_1, X_1<Y_1]=I(Y_1>X_1)\left\{\frac{1}{N^2}\left\{(N-2)[F(Y_1)-F(X_1)]+1\right\}\right.\\
&+\frac{n-m}{mN^2}\left\{\frac{N+1}{2}-(N-3)F(X_1)[1-F(X_1)]-F(Y_1)\right\}\\
&+\frac{m-n}{nN^2}\left\{\frac{N-1}{2}-(N-3)F(Y_1)[1-F(Y_1)]+F(X_1)\right\}\\
&\left.+\frac{mnN+6mn-2N^2}{2mnN^2} \left\{\frac{N-1}{3}+2F(X_1)[1-F(X_1)]+2F(Y_1)[1-F(Y_1)]\right\}\right\}.
\end{align*}
Hence 
\begin{align*}
&\E[T|X_1, Y_1, X_1<Y_1]\\
&=I(Y_1>X_1)\left\{\frac{mnN+N^2-7mn+5n-m}{6mnN}\right.\\
&+\frac{(2m-n)(n-1)}{mnN}F(X_1)[1-F(X_1)]
+\frac{(2n-m)(m-1)}{mnN}F(Y_1)[1-F(Y_1)]\\
&\left.+\frac{m-1}{mN}F(Y_1)-\frac{n-1}{nN}F(X_1)\right\}.
\end{align*}
A similar calculation gives
\begin{align*}
&\E[T|X_1, Y_1, X_1>Y_1]\\
&=I(Y_1<X_1)\left\{\frac{mnN+N^2-7mn+5m-n}{6mnN}\right.\\
&+\frac{(2m-n)(n-1)}{mnN}F(X_1)[1-F(X_1)]
+\frac{(2n-m)(m-1)}{mnN}F(Y_1)[1-F(Y_1)]\\
&\left.-\frac{m-1}{mN}F(Y_1)+\frac{n-1}{nN}F(X_1)\right\}.
\end{align*}
Therefore (\ref{conMxy}) holds. 

 Let $U_1, U_2$ be i.i.d. uniform random variables on $[0,1]$, then
\begin{align*}
&Var\{\E[T|X_1, Y_1]\\
&=Var\left\{\frac{(2m-n)(n-1)}{mnN}U_1(1-U_1)
+\frac{(2n-m)(m-1)}{mnN}U_2(1-U_2)\right.\\
&+I(U_2>U_1)\left\{\frac{5n-m}{6mnN}
+\frac{m-1}{mN}U_2-\frac{n-1}{nN}U_1\right\}\\
&\left.+I(U_2<U_1)\left\{\frac{5m-n}{6mnN}
-\frac{m-1}{mN}U_2+\frac{n-1}{nN}U_1
\right\}\right\}.
\end{align*}
Therefore (\ref{xyproj}) holds. 
This completes the proof.    \hfill{$\square$}

\vspace{4mm}


The following lemma gives the second order projection of $T$ on two $X$ variables or two $Y$ variables.
\begin{lemma}\label{tx1x2}
Under $H_0$, the projection of $T$ on two $X$ variables is 
\begin{align}
&\E[T|X_1, X_2]\nonumber\\
=&\frac{n}{N}\left\{\frac{mN+6n-m}{6mn}+\frac{m-2n}{mn}[F(X_1)(1-F(X_1))+F(X_2)(1-F(X_2))]\right.\nonumber\\
&\left.-\frac{1}{m}|F(X_2)-F(X_1)|\right\}.\label{projxy}
\end{align}
The variance of the projection is 
\begin{align}
Var \{\E[T|X_1, X_2]\}=\frac{m^2-2mn+5n^2}{90m^2N^2}.\label{xproj}
\end{align}
The projection of $T$ on two $Y$ variables is 
\begin{align*}
&\E[T|Y_1, Y_2]\\
=&\frac{m}{N}\left\{\frac{nN+6m-n}{6mn}+\frac{n-2m}{mn}[F(Y_1)(1-F(Y_1))+F(Y_2)(1-F(Y_2))]\right.\\
&\left.-\frac{1}{n}|F(Y_2)-F(Y_1)|\right\}.
\end{align*}
The variance of the projections is 
\begin{align}
&Var \{\E[T|Y_1, Y_2]\}=\frac{n^2-2mn+5m^2}{90n^2N^2}.\label{yproj}
\end{align}
\end{lemma}

\noindent{\bf Proof of Lemma \ref{tx1x2}}. By symmetry, we only need prove the results for the projection on two $X$ variables. By the definition of $T$ as in (\ref{eqn:statistic1}), under the null hypothesis,
\begin{align*}
&\E[T|X_1, X_2, X_1< X_2]\\
&=\frac{mn}{N}\left\{\frac{2}{m^2N}\E[|R_1-R_3||X_1, X_2, X_1< X_2]\right.\\
&+\frac{2}{m^2N}\E[|R_2-R_3||X_1, X_2, X_1< X_2]\\
&+\frac{mN-6n}{2m^2nN}\E[|R_3-R_4||X_1, X_2, X_1< X_2]\\
&\left.-\frac{1}{m^2N}\E[|R_1-R_2||X_1, X_2, X_1< X_2]\right\}.
\end{align*}
By (\ref{1212}), (\ref{1312}), (\ref{1312'}) and (\ref{3412}), we have
\begin{align*}
&\E[T|X_1, X_2, X_1< X_2]\\
=&\frac{n}{N}\left\{\frac{mN+6n-m}{6mn}+\frac{m-2n}{mn}[F(X_1)(1-F(X_1))+F(X_2)(1-F(X_2))]\right.\\
&\left.-\frac{1}{m}[F(X_2)-F(X_1)]\right\}I(X_1< X_2).
\end{align*}
By symmetry, we have (\ref{projxy}).
Let $U_1, U_2$ be i.i.d. uniform random variables on $[0,1]$, then
\begin{align*}
&Var \{\E[T|X_1, X_2]\}\nonumber\\
&=\frac{n^2}{N^2}Var\left\{\frac{m-2n}{mn}[U_1(1-U_1)+U_2(1-U_2)]-\frac{1}{m}|U_2-U_1|\right\}\nonumber\\
&=\frac{m^2-2mn+5n^2}{90m^2N^2}.
\end{align*}
This completes the proof.   \hfill{$\square$}

\vspace{4mm}

In the following we provide a lemma which is useful in deriving the asymptotics of the test statistic $T$.
  \begin{lemma}\label{ortho}
 Let $S_n(X_1, X_2, \cdots, X_n)$ be a function of $n$ independent  random variables with decomposition $S_n=M_n+R_n$. If  $\E(R_n)=Cov(M_n, R_n)=0$ for any $n$ and $Var(S_n)/Var(M_n)\rightarrow 1$ as $n\rightarrow \infty$, then
 $|R_n|/\sqrt{Var(S_n)}\rightarrow 0$ in $\mathcal{L}^2$ norm and therefore  $|R_n|/\sqrt{Var(S_n)}\rightarrow 0$ in probability.
 \end{lemma}
\noindent{\bf Proof of Lemma \ref{ortho}}. 
 \begin{align*}
 &\E [R_n^2/Var(S_n)]=\frac{\E [(S_n-\E S_n)-(M_n-\E M_n)]^2}{Var(S_n)}\\
&=\frac{Var(S_n)+Var(M_n)-2\E (S_n-\E S_n)(M_n-\E M_n)}{Var(S_n)}\\
&=\frac{Var(S_n)+Var(M_n)-2\E (M_n-\E M_n)^2-2\E R_n(M_n-\E M_n)}{Var(S_n)}\\
&=\frac{Var(S_n)-Var(M_n)}{Var(S_n)}=1-Var(M_n)/Var(S_n)\rightarrow 0.
 \end{align*} 
\hfill{$\square$}


The above Lemma \ref{ortho} is a result of H{\' a}jeck projection technique. See H{\' a}jek and {\v S}id{\' a}k \cite{Hajek67} and Hettmansperger and McKean \cite{Hettmansperger10} for details. 

\vspace{0.2cm}
\noindent{\bf Proof of Theorem \ref{proasym}}. We can write $h(x,y)=\sum_{k=1}^\infty\lambda_k\phi_k(x)\phi_k(y)$, where $\{\phi_k(\cdot)\}$ are the orthonormal eigenfunctions corresponding to the eigenvalues $\{\lambda_k\}$, see   Dunford and Schwartz  \cite{DunfordSchwartz},  Serfling \cite{Serfling}.
Since $\E h(X,y)=0$ and
$$0=Var \{\E [h(X,Y)|Y]\}=\sum_{k=1}^\infty \lambda_k^2(\E \phi_k(X))^2Var( \phi_k(Y)).$$
Therefore, $\E(\phi_k(X))=0$ for all $k\ge 1$ and $Var [h(X,Y)]=\E h^2(X,Y)=\sum_{k=1}^\infty\lambda_k^2$. By (\ref{var}),
$$\sum_{k=1}^\infty\lambda_k^2=Var [h(X,Y)]=2/45.$$
The above results can also be confirmed by $\lambda_k=-\frac{2}{\pi^2k^2}, \phi_k(x)=\cos k\pi F(x), k\in \mathbb{N}$. 
Denote $W_{mk}(X)=\frac{1}{\sqrt{m}}\sum_{i=1}^m\phi_k(X_i)$, $W_{nk}(Y)=\frac{1}{\sqrt{n}}\sum_{j=1}^n\phi_k(Y_j)$,
$Z_{mk}(X)=\frac{1}{m}\sum_{i=1}^m\phi_k^2(X_i)$, $Z_{nk}(Y)=\frac{1}{n}\sum_{j=1}^n\phi_k^2(Y_j)$.
Define $T_{NK}$ by 
\begin{align*}
&\sqrt{Var(T)}T_{NK}=\frac{1}{N}\sum_{i=1}^m\sum_{j=1}^n \sum_{k=1}^K\lambda_k\phi_k(X_i)\phi_k(Y_j)\nonumber\\
&-\frac{1}{N}\sum_{1\le i<j\le m}\sum_{k=1}^K\lambda_k\phi_k(X_i)\phi_k(X_j)-\frac{1}{N}\sum_{1\le i<j\le n}\sum_{k=1}^K\lambda_k\phi_k(Y_i)\phi_k(Y_j).\nonumber
\end{align*}
Then 
\begin{align}
&\sqrt{Var(T)}T_{NK}
=\frac{\sqrt{mn}}{N} \sum_{k=1}^K\lambda_kW_{mk}(X)W_{nk}(Y)\nonumber\\
&-\frac{m}{N}\sum_{k=1}^K\frac{\lambda_k}{2}\left\{W_{mk}^2(X)-Z_{mk}(X)\right\}
-\frac{n}{N}\sum_{k=1}^K\frac{\lambda_k}{2}\left\{W_{nk}^2(Y)-Z_{nk}(Y)\right\}.\label{TnK}
\end{align}
Applying the argument as in Serfling \cite{Serfling}, page 197,  $|\E e^{ix T_N}-\E e^{ixT_{NK}}|\le |x|[\E(T_N-T_{NK})^2]^{1/2}]$.
It is easy to see that
\begin{align*}
\E(T_N-T_{NK})^2&=\left(\frac{mn}{N^2}+\frac{m(m+1)+n(n-1)}{2N^2}\right)\frac{1}{Var(T)}\sum_{k=K+1}^\infty\lambda_k^2\\
&=[\frac{45}{2}+o(1)]\sum_{k=K+1}^\infty\lambda_k^2\le 23 \sum_{k=K+1}^\infty\lambda_k^2.
\end{align*}
For a given  $\epsilon>0$ and fixed $x$, we can choose and fix $K$ to be large enough so that
$$|x| (23 \sum_{k=K+1}^\infty\lambda_k^2)^{1/2}< \epsilon.$$
Then we have $|\E e^{ix T_N}-\E e^{ixT_{NK}}|<\epsilon$ for all $N$.
By (\ref{TnK}), Theorem \ref{t1var} and the condition $\lim_{N\rightarrow \infty}\frac{m}{n}=1$, we may write
\begin{align*}
T_{NK}
=\frac{\sqrt{45}}{2}\sum_{k=1}^K\lambda_k\left\{-[W_{nk}(X)-W_{nk}(Y)]^2/2+Z_{nk}(X)/2+Z_{nk}(Y)/2\right\}+r_N.
\end{align*}
It is easy to see  that  $r_N\rightarrow 0$ in probability. Let $\chi_{1k}^2$ be iid $\chi^2_1$ random variables. Denote $U_K=\frac{\sqrt{45}}{2}\sum_{k=1}^K\lambda_k(-\chi_{1k}^2+1)$ and $U=\frac{\sqrt{45}}{2}\sum_{k=1}^\infty\lambda_k(-\chi_{1k}^2+1)$. 
Since $\E W_{mk}(X)=\E W_{nk}(Y)=0$ and $W_{mk}(X), W_{nk}(Y)$ are orthonormal, then
the random vector $\{(W_{mk}(X)-W_{nk}(Y))/\sqrt{2}\}_{k=1}^K\Rightarrow N(0, I_{K\times K})$ by the Linderberg-Le\'{v}y central limit theorem, $Z_{mk}(X)\rightarrow 1$ and $Z_{nk}(Y)\rightarrow 1$ for $1\le k\le K$ by the strong law of large numbers.   Hence $T_{NK}\Rightarrow U_K$ as $n\rightarrow \infty$ and $\E(e^{ixT_{NK}})-\E(e^{ixU_K})|<\epsilon$ for all $N$. By the same argument as in \cite{Serfling},
$|\E(e^{ixU_K})-\E(e^{ixU})|<\epsilon$ for all $N$. Then together with $|\E e^{ix T_N}-\E e^{ixT_{NK}}|<\epsilon$, we have $|\E e^{ix T_N}-\E e^{ixU}|\rightarrow 0$ as $N\rightarrow \infty$.  Therefore,  $T_N\Rightarrow -\frac{\sqrt{45}}{2}\sum_{k=1}^\infty \lambda_k(\chi_{1k}^2-1)$,
where $\chi_{11}^2,\chi_{12}^2\cdots$ are independent $\chi_{1}^2$ variables.
Since $Var(\hat{T})/Var(T)\rightarrow 1$, we have $(T-\mathbb{E}T-\hat{\hat{T}})/\sqrt{Var(T)}\rightarrow 0$ in probability. Then  by Lemma \ref{ortho},
$$(T-\E T)/\sqrt{Var(T)}\Rightarrow -\frac{\sqrt{45}}{2}\sum_{k=1}^\infty \lambda_k(\chi_{1k}^2-1).$$ This completes the proof.  \hfill{$\square$}

\vspace{5mm}
\noindent{\bf Proof of Theorem \ref{thm:multi-inequality}}. Let $\mu$ be the uniform distribution on the surface of the unit ball $S^{d-1}=\{\bi a\in \mathbb R^d: \|\bi a\|=1\}$. From Theorem \ref{thm:inequality}, we have 
\begin{align}
&\E|R(\bi a^T\bi X,H^a)-R(\bi a^T\bi Y,H^a)|-\frac{1}{2}\E|R(\bi a^T\bi X_1,H^a)-R(\bi a^T\bi X_2,H^a)| \nonumber\\
&-\frac{1}{2}\E|R(\bi a^T\bi Y_1,H^a)-R(\bi a^T\bi Y_2,H^a)|\geq0 \nonumber
\end{align}
for each $\bi a\in S^{d-1}$,  where $H^a=\tau F^a+(1-\tau)G^a$ with $F^a$ and $G^a$ being the distributions of $\bi a^T\bi X$ and $\bi a^T\bi Y$ respectively. Integration of $\bi a$ with respect to $\mu$ obtains \eqref{eqn:mveqn}.  Equality holds if and only if for $\mu$-almost all $\bi a\in S^{d-1}$ the distributions of $\bi a^T \bi X$ and $\bi a^T \bi Y$ coincide.  For each $ t\in \mathbb R$ the functions $\E \exp(it \bi a^T\bi X)$ and $\E \exp(i t \bi a^T\bi Y)$ with $\bi a \in S^{d-1}$ are continuous. Thus, equality in \eqref{eqn:mveqn} holds if and only if $\bi X$ and $\bi Y$ have the same characteristic function, hence have the same distribution.  
\hfill{$\square$}

\end{document}